# Dispersive shock waves with nonlocal nonlinearity

Christopher Barsi, Wenjie Wan, Can Sun, and Jason W. Fleischer
*Department of Electrical Engineering, Princeton University, Princeton, New Jersey 08544*

**Abstract:** We consider dispersive optical shock waves in nonlocal nonlinear media. Experiments are performed using spatial beams in a thermal liquid cell, and results agree with a hydrodynamic theory of propagation.

Dispersion plays many roles in nonlinear optics, as the relative propagation of modes can either enhance or counteract the effects of mode coupling. Examples range from the balance of linear vs. nonlinear phase shifts in soliton formation[1] to radiative decay and supercontinuum generation[2]. Recently, the spatial equivalent of dispersion, i.e. diffraction, was used with defocusing nonlinearity to form dispersive shock waves in spatial beams[3]. Unlike dissipative shock waves, which have a thin, well-defined front due to energy absorption (e.g. viscosity), dispersive shock waves have a wavefront characterized by oscillations. In this case, dispersion and nonlinearity act in the same direction, creating a front profile resembling a soliton train. In hydrodynamic terms, wave breaking from nonlinear steepening is prevented by modal walk-off, resulting in a wave structure more commonly associated with dissipationless fluids, e.g. undular bores, collisionless plasma, and superfluids[4]. Here, we extend these studies to include the effects of a nonlocal nonlinearity. Experimentally and numerically, we find that nonlocality creates an effective damping force, inhibiting the shock speed and reducing oscillations by transferring momentum to larger scales.

Many optical materials respond nonlocally, in that the index of refraction at a particular location is determined by the intensity not only at that point, but at nearby portions of the material as well. The spatial extent of the index contribution at that point is assumed to be determined only by the properties of the medium itself. Nonlocal phenomena appear in many fields, such as plasma physics[5] and BEC[6], and can arise in optics through physical processes such as atomic diffusion[7] and thermal self-action[8,9]. Here, we use this latter effect and consider the self-defocusing of CW spatial beams in a thermal liquid cell.

The spatial beam dynamics considered here is modeled accurately by the nonlinear Schrödinger equation:

$$i\frac{\partial \psi}{\partial z} + \frac{1}{2k_0}\nabla_\perp^2 \psi + \Delta n(|\psi|^2)\psi = 0, \qquad (1)$$

where $\psi$ is the slowly-varying amplitude of the optical field, $k_0 = 2\pi n_0/\lambda$ is the wavenumber in a medium with base index $n_0$, and $\Delta n$ is the nonlinear index change induced by the intensity $\rho = |\psi|^2$. In many cases, a nonlocal nonlinearity can be modeled by a convolution operation: $\Delta n[\rho(\mathbf{r})] = \gamma \int R(\mathbf{r'})\rho(\mathbf{r-r'})d\mathbf{r'}$, where $R(\mathbf{r'})$ is the medium's response function normalized to unity, i.e., $\int R(\mathbf{r'})d\mathbf{r'} = 1$[10], and the integration takes place over the transverse dimensions of the system. For the thermal medium considered below, we assume a Gaussian response function $R(x) = (\pi w^2)^{-1/2}\exp(-x^2/w^2)$, where $w$ represents the effective range of the nonlocality.

To study shock behavior, it is helpful to express Eq. (1) in a more hydrodynamic form. This is done via the Madelung transformation $\psi(x,z) = \sqrt{\rho(x,z)}\exp[iS(x,z)]$[11], which, after scaling variables by $k_0$, yields two Euler-like fluid equations. In one dimension, these are

$$\frac{\partial \rho}{\partial z} + \frac{\partial}{\partial x}(\rho v) = 0, \qquad (2)$$

$$\frac{\partial v}{\partial z} + v\frac{\partial v}{\partial x} = \frac{\partial}{\partial x}\left[\Delta n'(|\psi|^2)\right] + \frac{1}{2}\frac{\partial}{\partial x}\left(\frac{1}{\sqrt{\rho}}\frac{\partial^2}{\partial x^2}\sqrt{\rho}\right), \qquad (3)$$

where $v = \partial S/\partial x$ is the fluid velocity and $\Delta n' = \Delta n/k_0$. Eq. (2) expresses conservation of power, while Eq. (3) is a momentum equation that states that optical flow results from nonlinear pressure and diffraction. We now consider a self-defocusing nonlinearity for which $\Delta n' < 0$, so that regions of high intensity cause outward flow. The convective derivative implies that light waves will undergo steepening as they propagate, ultimately leading to shock formation. For a symmetric response function, such as the Gaussian considered here, there is no "viscosity" (even derivative in Eq. 3, e.g. $\partial^2 v/\partial x^2$) in the system, so that wave breaking is prevented by the diffractive term in Eq. 3. This non-dissipative regularization leads to an oscillatory wavefront.

Numerical simulations of dispersive/diffractive shock waves in a nonlocal nonlinear medium are shown in Fig. 1. The input intensity profile consists of a Gaussian hump on a constant, low-level background, and output profiles are shown as the nonlocal parameter $w$ is increased. As in [3], the expanding wave consists of two repulsive humps whose fronts are characterized by oscillations. As the range of nonlocality increases, the central region broadens and the oscillations in the tails become damped. These effects are intuitively reasonable, because the convolution in $\Delta n$ should both broaden and smooth out the resultant field. Several limiting forms of the response function are useful to consider. In the limit of a delta-function response, the nonlinearity reduces to the local Kerr case $\Delta n = \gamma I$. (In this case, the background intensity defines an effective sound speed $c_s = \sqrt{\gamma \rho_0}$, so that the initial hump is automatically "supersonic."[3]) In the limit of a response width narrow compared to the input, the nonlinearity can be expanded in a Taylor series, giving $\Delta n = \gamma I + \gamma \left(\langle x^2 \rangle / 2\right)\left(d^2 I / dx^2\right)$, where $\langle x^2 \rangle = w^2$ is the average variance of the nonlocality. The second term has the same dispersive order as diffraction and has a spatially-dependent effect, weakening the effective repulsive pressure in the central part of the Gaussian beam while enhancing it in the tails. In the opposite limit of response width much greater than input beam width, the overall response (nonlinearity plus nonlocality) approaches the linear, local case; here, the response of the medium is equally strong across the entire beam profile, so that the hump does not experience a nonlinear phase shift relative to the background. Dynamically, though, higher-order terms and boundary effects arise in the longer-range limits, meaning that complex self-action and wave mixing are taking place[12,13].

Experimental measurements were made by projecting 532 nm laser light onto a 1 cm × 1 cm plastic cell containing ethanol doped with iodine. Physically, the iodine absorbs the green light and subsequently creates a thermal gradient in response to the light intensity. The heat diffuses throughout the cell and produces a nonlocal index of refraction via the thermo-optic effect: $\Delta n = (dn/dT)\Delta T$. Neglecting convection, the resulting system can be modeled by Eqs. 2-3. Following Ref. 3, we create a hump-on-background input profile by using a Mach-Zehnder interferometer and placing a spherical lens in one of the arms. Both beams are then recombined onto the input face of the cell, and the output face is imaged onto a CCD camera. In the liquid cell, the plane-wave background causes uniform heating and thus no relative index change, while the Gaussian hump produces a spatially-dependent nonlinearity. The liquid medium responds to this profile in two different stages: relatively fast thermal diffusion followed by convection of the fluid itself (after ~0.5s of CW heating). Measuring the output profile at different times allowed different stages of spreading (and thus nonlocality) to be observed. Fig. 2(a) shows the output for linear diffraction. The small oscillations result from the interference between the background and the curved Gaussian wavefronts. Fig. 2(b) shows the initial stage of the nonlocal shock evolution. The thermal gradient is weak, but not negligible, so that the beam diameter increases dramatically to about five times the diameter in the linear case. The next panel, Fig. 2(c) shows the full beam expansion due to the increased temperature gradient and hence nonlocality. For longer times, the shock begins to move vertically and develop an asymmetry [Fig. 2(d)], due to convection of the liquid medium itself.

As the hump-to-background ratio increases, the shocks become more and more violent, with faster oscillations (higher self-phase modulation) in the tails. Taking the shock width as a measure of its velocity (Fig. 3), it is found that the front speed increases with intensity ratio according to the relation $D_s = a_s\left(1 + b_s \sqrt{\rho/\rho_\infty}\right)$, with fitting parameters $a_s = 185\ um$ and $b_s = 0.13$. This is the same scaling as obtained in Ref. 3, but with a rate of increase much slower than in the local, Kerr case (Fig. 3, inset). For the liquid medium here, the $b_s$-coefficient suggests a nonlocal response of roughly twice the width of the input hump.

Figure 4 shows the effects of the background intensity, obtained by measuring the output for different power levels both with (top row) and without (bottom row) the background. Each column shows the output profiles for input Gaussian powers of 6.1 mW, 24.4 mW, and 36.6 mW, respectively. Interestingly, even without the background, oscillations are still apparent in the output. Known as thermal blooming[8, 9, 14, 15], this single-beam broadening has recently been interpreted in a shock context[16]. The shock waves observed here, however, differ from these more classical defocused waves in two ways: the central region is broad and flat, without an inner series of concentric circles, and the outer rings have a decreasing period with radial distance. This latter characteristic is a signature feature of dispersive shock waves[4] and results in our case from a steep initial gradient, forcing the initial Gaussian hump to wave-break into its own tails.

In conclusion, dispersive optical shock waves have been characterized numerically and experimentally in a nonlinear, nonlocal medium. A hydrodynamic model was developed, showing that shock formation is inhibited by the nonlocal response. It was shown that nonlocality acts as a nonviscous damping force, reducing the shock speed and inhibiting oscillations in the front. The results extend traditional laser observations of thermal blooming and reinforce the potential of optical systems to model hydrodynamic flow.

**List of Figures**

**Fig. 1.** Numerical simulation of dispersive shock waves as a function of nonlocal response width ($w'$, normalized to the width of the input Gaussian hump). The red bars on the x-axis indicate the FWHM of the initial hump, which has a 25:1 peak-to-background intensity ratio.

**Fig. 2.** Experimental evolution of diffractive shock wave after 1cm of propagation in an ethanol + iodine liquid cell. (a) Linear case. Inset: input profile. (b) Initial shock formation in nonlinear, nonlocal case. (c) ~200 ms later, quasi-steady state (before convection of fluid) shock profile. (d) Steady state, with asymmetry due to convection.

**Fig. 3.** Shock length, measured horizontally from centerline to the end of oscillations, as a function of hump-to-background intensity ratio. Black dots: experimental measurements. Solid line: best fit of $D_s = a_s\left(1 + b_s\sqrt{\rho/\rho_\infty}\right)$, with $a_{nl}$ = *185 um* and $b_n$ = *0.13*. Inset: numerical simulation, showing consistency of scaling relation as a function of nonlocal response width $w'$.

**Fig. 4.** Comparison of shock profiles with background (a-c) and without background (d-f). From left to right, the laser power is 6.1 mW, 24.4 mW, and 36.6 mW, respectively.

**Fig. 1**

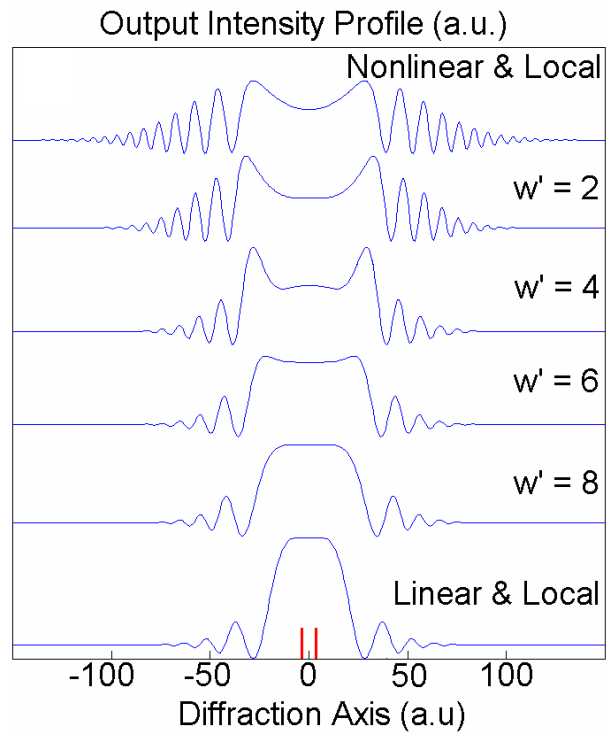

**Fig. 2**

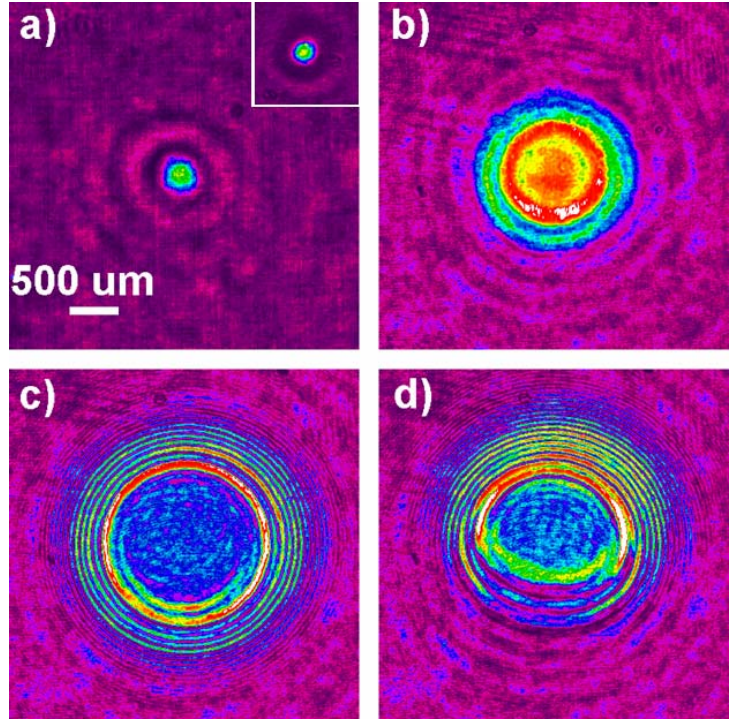

**Fig. 3**

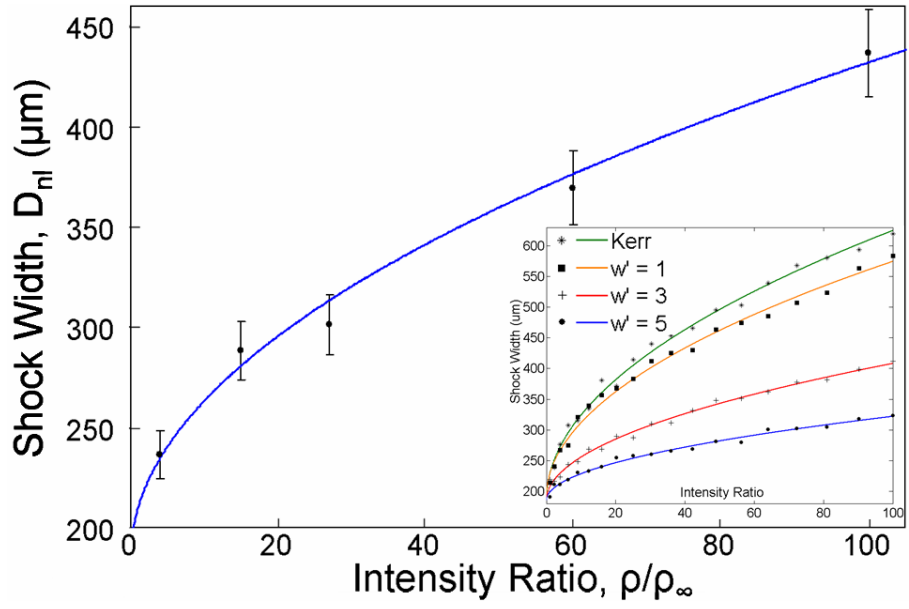

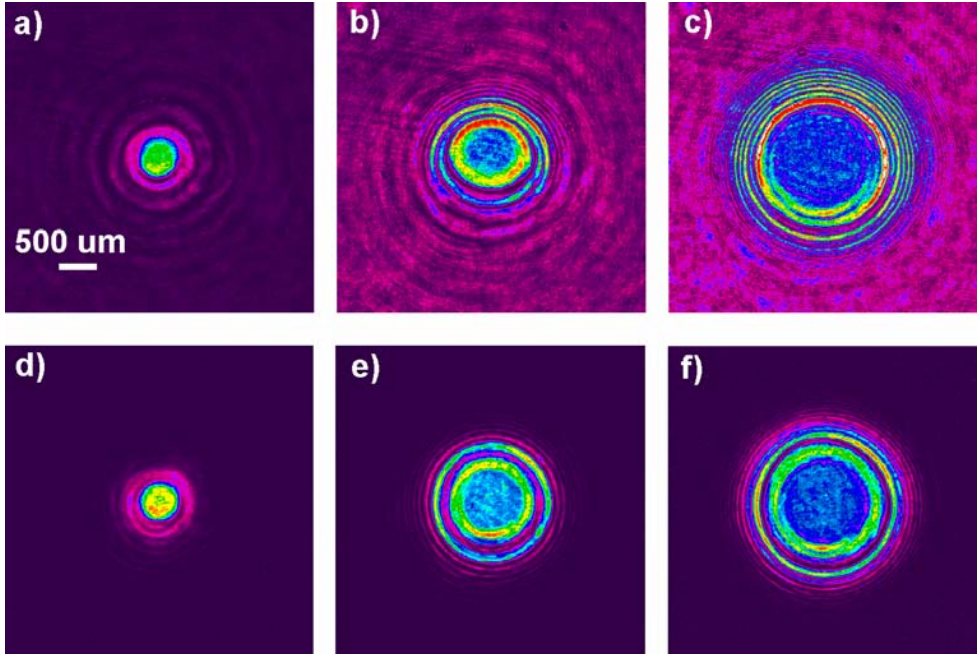